\documentclass[12pt]{iopart}

\usepackage{iopams}  
\usepackage{graphicx}
\begin{document}

\title[Crucial inputs to nucleosynthesis calculations]{Crucial
inputs to nucleosynthesis calculations}

\author{T Rauscher}

\address{
%$^1$
Dept.\ of Physics and Astronomy, University of Basel, Basel, Switzerland
}
%\address{$^2$ Atomic Institute of the Austrian Universities, Vienna, Austria}
\ead{Thomas.Rauscher@unibas.ch}
\begin{abstract}
The first part of the paper discusses nuclear properties relevant to predict
compound reactions. The second part addresses direct reactions with special
emphasis on direct neutron capture.
\end{abstract}

%Uncomment for PACS numbers title message
%\pacs{00.00, 20.00, 42.10}
% Keywords required only for MST, PB, PMB, PM, JOA, JOB? 
%\vspace{2pc}
%\noindent{\it Keywords}: Article preparation, IOP journals
% Uncomment for Submitted to journal title message
%\submitto{\JPA}
% Comment out if separate title page not required
%\maketitle

\section{Introduction}
Reaction rates are central in the modeling of nucleosynthesis
and astrophysical energy generation and are computed by an integration of
reaction cross sections across an energy range defined by the plasma
temperature. Due to the comparatively low energies
($E\lesssim 10$ MeV for charged projectiles and $E\lesssim 1$ MeV for
neutrons)
relevant in astrophysical plasmas, mainly three reaction mechanisms are
important \cite{descrau}: compound, resonant, and direct reactions.
In principle, both resonant and direct
mechanism may contribute to the reaction cross section although one of them
is negligible in most cases. The nuclear level density in the compound
nucleus, at the energy at which the compound nucleus would be formed,
determines which mechanism is dominating \cite{rtk97}. The statistical
Hauser-Feshbach model can be applied when there are many, overlapping
resonances which can be treated with averaged transmission coefficients.
In the absence of resonances and strong tails of resonances, the direct
mechanism will be important even at low energy \cite{descrau}. In between the
two regimes, the contributions of single resonances require the precise
knowledge of their properties, including energy, spin, parity, as well as
the phases of wave functions to account for possible interference effects.

Describing the interplay of different mechanisms has been a longstanding
problem in the study of nuclear reactions and requires precise predictions
of excited state properties. A thorough discussion is beyond the scope of
this paper but it is emphasized that this, among other things,
shows the great importance
of measuring cross sections {\it in the relevant energy range}.
Here, we are focusing on some of the
ingredients necessary for calculations in the Hauser-Feshbach model.
One of the main difficulties in the determination of
reaction rates for astrophysics lies in the fact that most reactions
involve unstable nuclei which are not (yet) accessible in the laboratory.
Therefore, the required properties cannot be extracted from experiment
but rather have to be predicted by theory. In the second part of the
paper we discuss direct capture and compare the
results to Hauser-Feshbach cross sections for selected nuclides.

\section{Compound reactions}

\subsection{Important nuclear properties}

Averaged widths (or transmission functions) are the central quantities
in the calculation of cross sections in the Hauser-Feshbach model \cite{hf}.
The relevance of the nuclear level density not only for the
identification of the reaction mechanism but also for the computation of
transmission functions has been discussed elsewhere \cite{descrau,rtk97,moc}.
The sensitivity of the results to variations in the level density is
not strong because
$i$) transitions to low lying states are dominating and these are
usually explicitly included \cite{adndt1,adndt2}, and $ii$) the thermal population of target
states in the plasma allows for additional transitions washing out the
effect of non-equally distributed parities \cite{moc}.

Averaged radiative widths are usually calculated including E1 and M1 transitions \cite{nonsmoker,adndt1,most}, more recently also E2 \cite{websmoker}.
For E1 transitions, Lorentzian shapes with modified low-energy tails
\cite{adndt1} or strength functions from QRPA calculations \cite{gorkhan}
are used. Frequently, the appearance of low-lying additional GDR strength
is discussed, e.g.\ stemming from soft-mode vibrations, often termed ``pygmy
resonance''. This could, in principle, strongly enhance neutron capture
cross sections of very neutron-rich nuclei \cite{most,gorkhan}. However,
it will only be important for r-process nucleosynthesis if three conditions
are met: $i$) the additional strength has to be located below the neutron
separation energy of the nucleus because otherwise there will be no
$\gamma$ transitions affected, $ii$) the statistical model has to be
applicable because otherwise direct capture may be dominating which does
not see the GDR+pygmy strength, and $iii$) the involved nuclei should not
participate in reaction equilibria ((n,$\gamma$)-($\gamma$,n) equilibrium)
but rather be produced by individual captures or photodisintegrations.
The last item calls for nuclei located outside the main r-process path
which are reached in the brief freeze-out phase.
There is no consistent picture yet.
Previous microscopic calculations found some
strength below the neutron threshold \cite{tso}, while
recent calculations find the pygmy resonance consistently several MeV above the
neutron separation energy \cite{vret1,vret2,vretthis,piek06,liang}.
The latter would rule out an
effect on astrophysical rates provided the width is not so large as to
reach below the separation energy.
The location of the additional strength above the separation energy
for Sn isotopes
is confirmed by a recent experiment \cite{gsi1,gsi2,gsithis}.
Earlier experiments found additional E1 strength below the separation
energy in semi-magic isotopes of Ba, Ce, and Sm \cite{zilges}.
Future experiments have to
test the predictions further away from stability.

Another complication in the prediction of $\gamma$ transitions arises
due to isospin selection rules suppressing E1 and M1 $T=0 \rightarrow T=0$
transitions. This leads to a strong suppression of ($\alpha$,$\gamma$)
cross sections of self-conjugate nuclei \cite{rau2000} and some suppression
of (p,$\gamma$) and (n,$\gamma$) on such nuclei. Often, this
effect is included by introducing arbitrary suppression factors \cite{most,woo}
because the original Hauser-Feshbach equation implicitly assumes complete
isospin mixing \cite{hf}. It is possible to generalize the model to
explicitly treat isospin mixing \cite{nonsmoker,rau2000}.

Nuclear masses determine the reaction $Q$-values in the open channels.
Beyond the region of measured masses, theoretical models have to be employed.
Special attention has to be paid to the transition from the measured to
the unmeasured region when computing $Q$-values. Special care has to be
taken to avoid artificial breaks in the relative channel energies when
subtracting values from different sources.

\subsection{Optical potentials}
Transmission functions for particles
are usually predicted by utilizing the optical model \cite{descrau}.
This requires an optical potential with real and imaginary parts that are,
in principle, dependent on the type of the projectile,
target mass, and projectile energy. Because of the sensitivity of the
transmission coefficients to the depth and geometry of a potential,
the calculated cross sections are also very sensitive to the choice of
potential. Global potentials are necessary for large-scale predictions which
should also give reliable results off stability where no experimental
information is available.

The largest uncertainties in this respect are in optical potentials for charged
projectiles due to the low incident energies occurring in astrophysical
applications. The relevant energies are close to the Coulomb barrier. This
makes it extremely difficult not just to measure cross sections but also
to determine optical potentials by scattering \cite{mohr144}. Available
global potentials are usually derived at higher energies and/or only for a
limited mass range.

The microscopic optical potential of \cite{jlm} (JLM)
with low-energy modifications
by \cite{jeuk} is widely used for neutrons and protons.
With a few exceptions it has been successful in reproducing experimental
reaction data across a large range of target masses. Recently, new
parameterizations of the JLM potential have been derived \cite{baugefirst,bauge}
with the aim to improve its performance at energies above 160 MeV. However,
this new potential does not work at low energies, as has been found in
several comparisons to recent data \cite{gyu03,kiss}, whereas the JLM
potential performed much better. However, it has been argued that the
isovector components of the JLM potential may be too weak \cite{baugefirst}.
Indeed, in a comparison to recent $^{70}$Ge(p,$\gamma$) and
$^{76}$Ge(p,n) data \cite{kiss}
we find that the absolute value and the energy-dependence of the
$S$-factors of this reaction and of previously measured (p,$\gamma$) and
(p,n) reactions on Se and Sr isotopes can be better reproduced when
increasing the JLM imaginary depth by 70\%. This can be attributed to a
stronger isovector component in the imaginary part. For further details,
see \cite{kiss}.

Regarding optical potentials for $\alpha$ particles, see \cite{demetthis}
for an extended discussion. There are no microscopic potentials available
and attempts to derive a global potential are restricted to fits of
Woods-Saxon shapes. Interestingly, despite of the higher Coulomb
barrier, $\alpha$ captures at low energies can still be described within
a factor of two by the simple potential of \cite{satch} which was derived
from $\alpha$ scattering at energies of several tens of MeV. However, a
trend is found in measured values which often are well predicted theoretically
closely above the Coulomb barrier height and exhibiting a slowly widening
difference at sub-barrier energies (see, e.g., \cite{ozag,bas07}).
This indicates that the barrier transmission may not be described
correctly and that an additional dependence on barrier height
may have to be introduced in the potentials. Another problem may arise
due to the formation of $\alpha$ clusters in the compound nucleus.
The difficulty in describing low-energy $\alpha$-induced reactions is
similar to the well-known problem of predicting half-lives of
spontaneous $\alpha$-emitters.

There is one remarkable case, $^{144}$Sm($\alpha$,$\gamma$),
showing the by far strongest discrepancy between experiment and theory
at the lowest measured energy of 10.2 MeV \cite{som}. Because of the importance
of the extrapolation to 9 MeV for $\gamma$-process nucleosynthesis, it
is urgently required that this reaction be remeasured at sub-Coulomb
energies!

In general, when performing measurements aiming at an improvement in
optical potentials one has to select reactions properly. The potential
of interest should give rise to the smallest width of the involved
channels because otherwise the sensitivity of the cross section to a
variation of the potential is small \cite{kiss,gyurkyag}. For $\alpha$ potentials this can
be achieved by studying ($\alpha$,$\gamma$) around the Coulomb barrier
or in ($\alpha$,n) and (n,$\alpha$) reactions. For the proton potential,
(p,n) or (n,p) reactions can be useful. This has also to be considered
when fitting optical potentials to reaction data. Only reactions
sensitive to the optical potential will provide useful constraints.
Such fits including reaction data may be of particular importance because
it has been pointed out recently that there may be a fundamental
difference between potentials derived from scattering on a ``cool''
target and emission from a ``hot'' nucleus \cite{avri}. A dependence on
nuclear temperature may have to be included in the potentials.

\section{Direct capture}

\subsection{Importance for neutron-rich nuclei}
The vast majority of reactions relevant for astrophysics can be treated in
the statistical model. However, the nuclear level density is low in
nuclei which are light ($A\lesssim 20$) or have closed shells.
Towards the driplines, the neutron- or proton-separation energies are decreasing which
leads to formation of the compound nucleus at low excitation energy and
thus with a low level density at that formation energy. In all these cases,
the Hauser-Feshbach model will not be applicable anymore and has to be
supplemented by accounting for resonant and direct processes. This is
straightforward provided the required nuclear properties are known and the
mechanisms are clearly separated. The latter will be the case when one of
the reaction mechanisms dominates. If two of the mechanisms exhibit similar
strength, optical potentials may have to be modified to account for the
increased flux into the other reaction channel. Additionally, coherent
summation of the channels may be required to allow for interference
effects.

\begin{figure}
\begin{center}
\includegraphics[angle=-90,width=\linewidth]{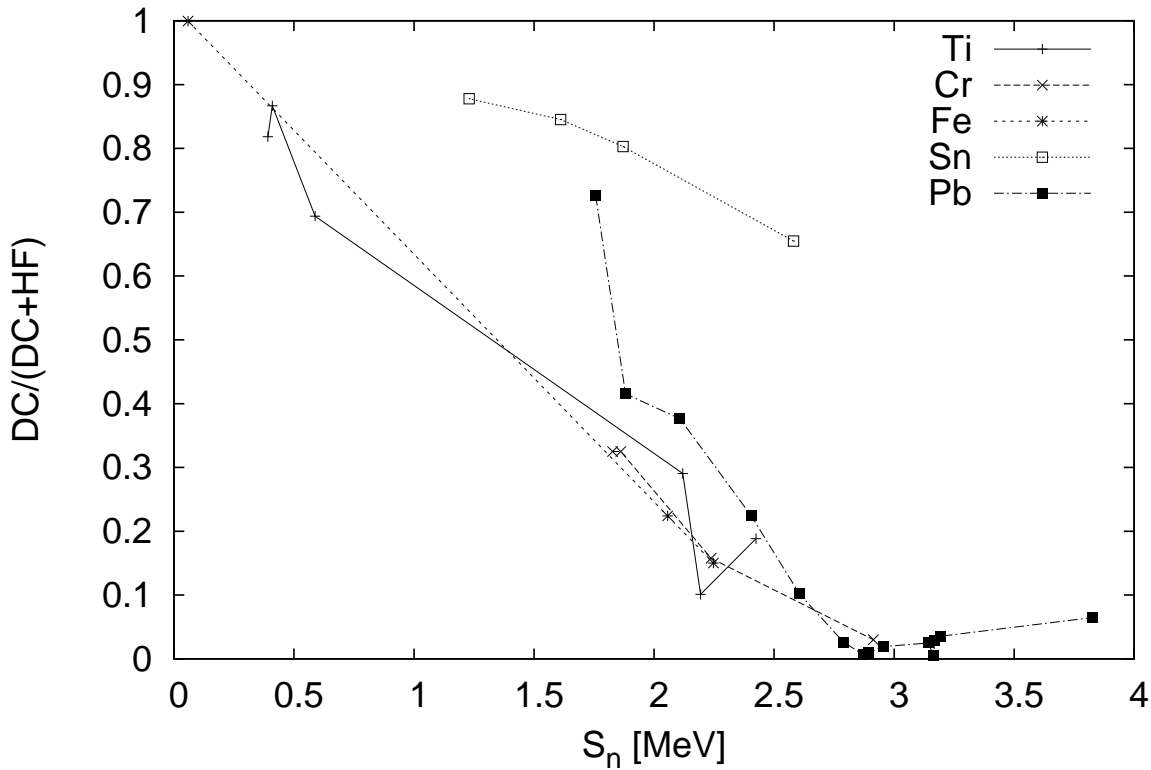}
\caption{\label{fig:dc_sep}Relation between direct neutron capture and compound capture as
function of neutron separation energy for isotopes of Ti, Cr, Fe \cite{enam95},
Sn \cite{budapest94}, and Pb \cite{raudc} }
\end{center}
\end{figure}

Even in heavy, closed shell nuclei direct capture may have a non-negligible
contribution as, e.g., was demonstrated in \cite{raudc} where a discrepancy
between an activation and a resonance measurement of $^{208}$Pb(n,$\gamma$)
was shown to be caused by this mechanism. It becomes dominant
in the absence of resonances. More specifically, it will be non-negligible
for targets close to the driplines. Such nuclides are produced in the rp-process
on the proton-rich side and the r-process at the neutron-rich side of the nuclear chart.
Figure \ref{fig:dc_sep} shows the increasing importance of direct neutron
capture with decreasing neutron separation energies. (It is to be noted that these
estimates just compare compound and direct cross section without accounting for
interference effects.)

\subsection{Statistical direct capture}
It has been shown \cite{raudc} that the direct cross section is very sensitive to the
predicted properties of the final states and that different microscopic
models yield vastly different results. The
sensitivity is much higher than in the case of compound reactions
because no averaged quantities are used. In order to
circumvent the problem of the exact prediction of states it
has been suggested \cite{raurep96,holz97,gordc97} to employ averaged
properties also for direct captures, i.e.\ to replace the sum over discrete
final states by an integration over a level density. Determination of the nuclear
level density poses the same problem as in the case of compound reactions.
Additionally, spectroscopic factors have to be predicted. Previously, these were
assumed to be constant \cite{gordc97} or energy-dependent \cite{raurep96,holz97,ejn98}.

\begin{figure}
\begin{center}
\includegraphics[width=\linewidth]{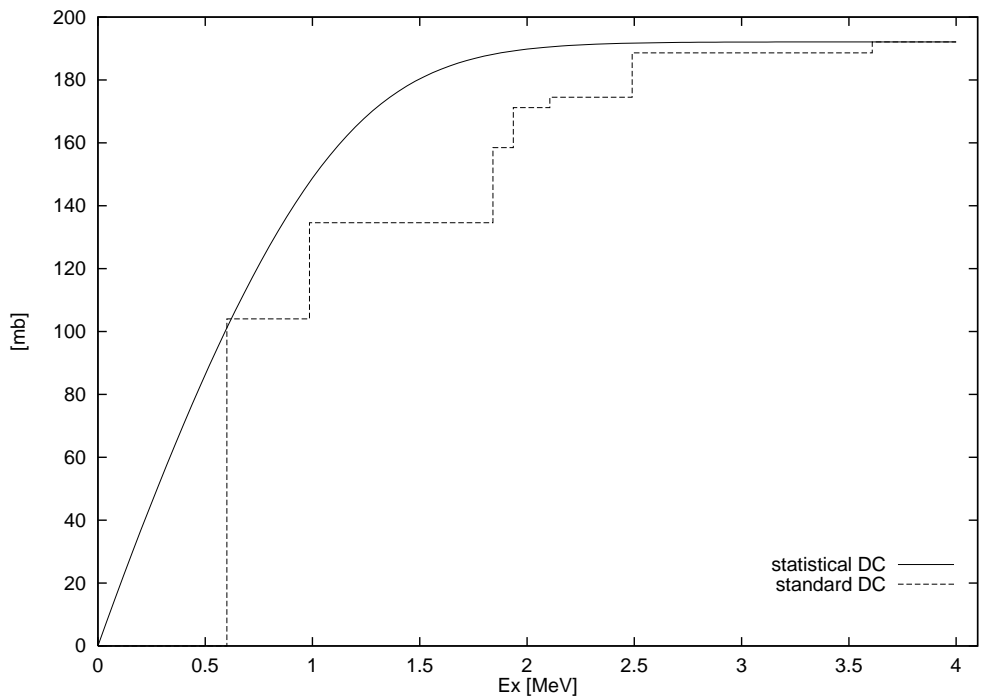}
\caption{\label{fig:avdc}Sums of the contributions to the total direct capture
cross sections with the averaged (statistical) DC model and the standard DC
for $^{136}$Xe(n,$\gamma$) as function of excitation energy \cite{holz97} }
\end{center}
\end{figure}

\begin{figure}
\begin{center}
\includegraphics[width=\linewidth]{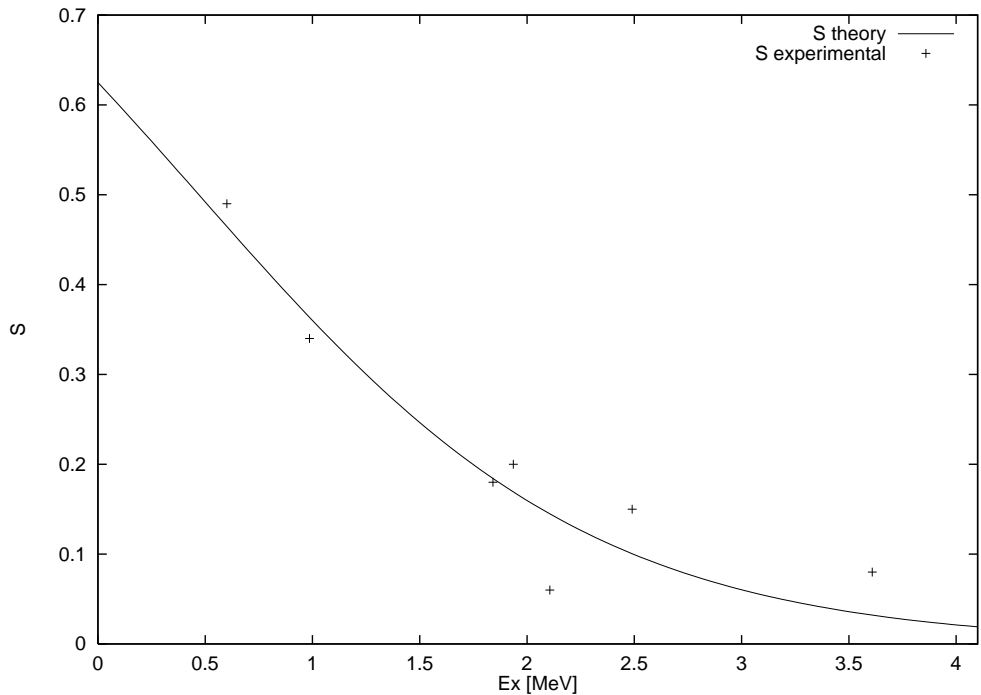}
\caption{\label{fig:s}Comparison between experimental and averaged (equation \ref{eq:s})
spectroscopic factors
as function of excitation energy for $^{136}$Xe+n
\cite{holz97} }
\end{center}
\end{figure}

The spectroscopic factors describe the overlap between the antisymmetrized wave
functions of target+nucleon and the final state.
The number of final state configurations increases with increasing excitation energy
and the overlap of initial and final state wavefunctions decreases. Thus, also the
spectroscopic factor $S$ decreases. In a simple approach, the energy dependence of the
spectroscopic factor can be parameterized by a Fermi function with
\begin{equation}
\label{eq:s}
S=\frac{1}{1+e^\frac{E_{\rm x}-E^*}{\Delta}}
\end{equation}
and the parameters $E^*,\Delta$.
Despite of its simplicity, direct capture cross sections can be well described by
this approach \cite{holz97}. Figure \ref{fig:avdc} shows a comparison between
averaged direct neutron capture and standard neutron capture calculated in the Lane-Lynn
model \cite{lynn} for $^{136}$Xe and using the spectroscopic factors shown in Figure \ref{fig:s}.
For one-nucleon capture,
$S$ can also be derived
from the occupation probability $v^2$ for the target state into which the nucleon
is captured. It can be
calculated, e.g., by employing BCS \cite{moc} or Lipkin-Nogami pairing \cite{enam95,raudc}.
For the case of one-nucleon
capture on even-even nuclei, the spectroscopic factor $S$ for capture can be reduced to
$S=1-v^2$ \cite{enam95,glen}.

Recently, it was found that thermal excitation of target nuclei reduces the
sensitivity to the parity dependence in the nuclear level density \cite{moc}. We can
expect a similar effect for direct capture, although the number of possible transitions is
more limited \cite{moc}. A detailed discussion will be presented in a forthcoming paper
\cite{raufuture},
along with large-scale
calculations in the averaged direct capture model with excitation-energy dependent level densities and spectroscopic factors.

\section{Summary and conclusion}

The sensitivities of reaction cross sections to predicted nuclear properties
were discussed here for compound and direct reactions. Obviously, strong single
resonances in the relevant energy range will also have considerable impact on the
resulting astrophysical reaction rates. Because the resonance energies have to be
predicted to a few hundred keV, this poses a continuous challenge to nuclear theory.
For a detailed study of the actual impact of rate variations on nucleosynthesis in massive
stars for nuclides with $28\leq A\leq 80$, see \cite{hoffman}.

\end{document}